\documentclass[11pt]{article}

\usepackage{amsmath}
\usepackage{amssymb}
\usepackage{booktabs}
\usepackage{graphicx}
\usepackage[colorlinks=true,linkcolor=blue,citecolor=blue,urlcolor=blue]{hyperref}
\usepackage[round]{natbib}
\usepackage[margin=1in]{geometry}
\usepackage{xcolor}

\title{Association $\neq$ Similarity: Learning Corpus-Specific\\Associations for Multi-Hop Retrieval}
\author{Jason Dury\\Independent Researcher\\jason@eridos.ai}
\date{}

\begin{document}

\maketitle

\begin{abstract}
\noindent Dense retrieval systems rank passages by embedding similarity to a query, but multi-hop questions require passages that are \emph{associatively related} through shared reasoning chains. We introduce Association-Augmented Retrieval (AAR), a lightweight transductive reranking method that trains a small MLP (4.2M parameters) to learn associative relationships between passages in embedding space using contrastive learning on co-occurrence annotations. At inference time, AAR reranks an initial dense retrieval candidate set using bi-directional association scoring. On HotpotQA, AAR improves passage Recall@5 from 0.831 to 0.916 (+8.6 points) without evaluation-set tuning, with gains concentrated on hard questions where the dense baseline fails (+28.5 points). On MuSiQue, AAR achieves +10.1 points in the transductive setting. An inductive model trained on training-split associations and evaluated on unseen validation associations shows no significant improvement, suggesting that the method captures corpus-specific co-occurrences rather than transferable patterns. Ablation studies support this interpretation: training on semantically similar but non-associated passage pairs degrades retrieval below the baseline, while shuffling association pairs causes severe degradation. A downstream QA evaluation shows retrieval gains translate to +6.4 exact match improvement. The method adds 3.7ms per query, trains in under two minutes on a single GPU, and requires no LLM-based indexing.
\end{abstract}

\section{Introduction}
\label{sec:introduction}

Consider the question: \emph{``What is the birthplace of the director of Pulp Fiction?''} A retrieval system must find a passage about Quentin Tarantino directing Pulp Fiction and a separate passage about Tarantino's birthplace. The first passage ranks highly because it shares entities with the query. The second, about a person born in Knoxville, Tennessee, bears little surface similarity to a question about film directing. Yet both passages are required for a correct answer.

This failure pattern is systematic in multi-hop question answering. Dense retrievers find passages that resemble the query but miss passages that are needed alongside other retrieved passages to complete a reasoning chain. For multi-hop QA, relevance to the query and relevance to other supporting passages are distinct signals; dense retrieval handles the first well but often misses the second.

The Predictive Associative Memory (PAM) framework \citep{dury2026pam} provides formal grounding for this distinction, arguing that cosine similarity and learned associative retrieval produce different retrieval rankings. PAM proposes that association recovers items that were experientially co-present, regardless of their perceptual similarity, and predicts that such retrieval should be specific to experienced co-occurrences.

We operationalise this insight as Association-Augmented Retrieval (AAR), a transductive method that learns passage-to-passage associations from co-occurrence annotations and uses them to rerank dense retrieval results. The approach is deliberately minimal: a 4-layer MLP trained with contrastive loss on passage pairs that co-occur as supporting facts for the same question. At inference time, bi-directional association scoring blends learned association strength with cosine similarity to rerank an initial candidate set.

AAR learns associations over the target corpus and is evaluated on questions drawn from that corpus. This mirrors how RAG systems are typically deployed: the retrieval index and any auxiliary structures are built for a specific document collection. We show that this transductive approach improves multi-hop retrieval on both benchmarks tested, while an inductive variant (trained on training-split associations only) does not, providing empirical support for corpus-specific association learning.

Our contributions:
\begin{enumerate}
    \item Empirically, a lightweight transductive association function improves multi-hop passage retrieval by +8.6 Recall@5 on HotpotQA (without evaluation-set tuning) and +10.1 on MuSiQue, with gains concentrated on questions where dense retrieval fails (+28.5 points on hard questions).
    \item Ablations indicate that association and similarity produce opposite effects in multi-hop retrieval: training on semantically similar but non-associated pairs \emph{degrades} performance, while shuffling associative pairs causes severe degradation---both on HotpotQA.
    \item An inductive model trained on training-split associations shows no significant improvement on either dataset, pointing to corpus-specific co-occurrence learning.
    \item Operationally, the method requires 4.2M parameters, two minutes of training, and 3.7ms per-query overhead, with no LLM-based indexing.
\end{enumerate}

\section{Related Work}
\label{sec:related}

\subsection{Dense Retrieval for RAG}

Retrieval-augmented generation \citep{lewis2020rag} conditions language model outputs on retrieved passages, reducing hallucination and enabling knowledge-intensive tasks. Dense Passage Retrieval \citep[DPR;][]{karpukhin2020dpr} established the paradigm of encoding queries and passages into a shared embedding space and retrieving by inner product. Modern embedding models such as BGE \citep{xiao2024bge} improve retrieval quality through instruction-tuned contrastive training but remain similarity-based.

\subsection{Reranking Approaches}

Cross-encoder rerankers score query--passage pairs jointly and improve precision over bi-encoder retrieval, but at significant computational cost since every candidate requires a full forward pass. Learned rerankers typically optimise for relevance to the query. AAR scores \emph{inter-passage association} instead, capturing a complementary signal.

\subsection{Graph-Augmented RAG}

GraphRAG \citep{edge2024graphrag} constructs knowledge graphs from corpora using LLM extraction, enabling structured traversal for multi-hop queries. RAPTOR \citep{sarthi2024raptor} builds hierarchical document trees through recursive summarisation. These approaches improve multi-hop retrieval but require extensive LLM-based preprocessing: GraphRAG processes every passage through an entity and relationship extraction pipeline, incurring millions of LLM tokens for corpus indexing. AAR pursues comparable goals at lower computational cost, though direct performance comparison is not possible due to differences in experimental setup. We compare computational cost and architectural complexity only; controlled performance comparisons would require matched conditions.

\subsection{Neurobiologically-Inspired Approaches}

HippoRAG \citep{gutierrez2024hipporag} draws on hippocampal indexing theory to construct a knowledge graph where passages are linked through shared entities extracted by an LLM. HippoRAG~2 \citep{gutierrez2025hipporag2} extends this with improved entity linking and retrieval mechanisms. Both HippoRAG and AAR learn passage-to-passage relationships, but through different mechanisms: HippoRAG builds an explicit knowledge graph via LLM entity extraction, while AAR learns an implicit association function in embedding space. Both are transductive.

\subsection{Predictive Associative Memory}

The PAM framework \citep{dury2026pam} formalises the distinction between similarity-based and association-based retrieval using JEPA-inspired \citep{lecun2022path} predictive architectures. PAM argues that a predictor trained on co-occurrence patterns retrieves items that cosine similarity misses, and predicts that such retrieval should be specific to experienced associations. The present work tests both predictions on standard multi-hop QA benchmarks: transductive association learning improves retrieval on both datasets, while inductive transfer fails.

\section{Method}
\label{sec:method}

\subsection{Problem Formulation}
\label{sec:problem}

Let $\mathcal{C} = \{p_1, \ldots, p_N\}$ be a corpus of $N$ passages and $e: \mathcal{C} \to \mathbb{R}^d$ an embedding function mapping passages to unit vectors. Standard dense retrieval ranks passages by cosine similarity to a query $q$:
\begin{equation}
\text{sim}(q, p_i) = e(q)^\top e(p_i)
\label{eq:cosine}
\end{equation}
Multi-hop questions require passage sets $\{p_a, p_b\}$ where $p_a$ may be similar to $q$ but $p_b$ is connected to $p_a$ through a reasoning chain rather than to $q$ directly.

We define an \emph{association} between passages $p_a$ and $p_b$ as a relationship indicating that both are required to answer the same question---they co-occur as supporting facts. Associated passages need not be similar: a passage about a film director and a passage about a city's demographics may be associated (through the director's birthplace) while being distant in embedding space.

\subsection{Association Model}
\label{sec:model}

We train a function $f: \mathbb{R}^d \to \mathbb{R}^d$ to map passage embeddings into an \emph{association space} where associated passages are close and unassociated passages are distant. The architecture is a 4-layer MLP with LayerNorm, GELU activations, and a learned residual connection:
\begin{equation}
f(x) = \text{normalize}\bigl(\alpha \cdot x + (1 - \alpha) \cdot g(x)\bigr)
\label{eq:residual}
\end{equation}
where $g$ is the MLP transformation, $\alpha$ is a learned scalar blending the input with the transformed output, and the result is L2-normalised to lie on the unit hypersphere. The hidden dimension matches the input dimension (1024), yielding 4,204,545 parameters.

The residual connection preserves the original embedding's information while learning an associative perturbation. The learned $\alpha$ controls how far the output deviates from the input.

\subsection{Training}
\label{sec:training}

Given a set of association pairs $\mathcal{A} = \{(p_a^{(i)}, p_b^{(i)})\}$ derived from co-occurrence annotations (passages that serve as supporting facts for the same question), we train $f$ with a symmetric contrastive loss \citep{radford2021clip}. For a batch of $B$ pairs, we compute:
\begin{equation}
s_{ij} = f(e(p_a^{(i)}))^\top f(e(p_b^{(j)})) \,/\, \tau
\label{eq:sim_matrix}
\end{equation}
where $\tau$ is a temperature parameter. The loss is the average of row-wise and column-wise cross-entropy:
\begin{equation}
\mathcal{L} = \frac{1}{2}\left[\text{CE}(\mathbf{S}, \mathbf{y}) + \text{CE}(\mathbf{S}^\top, \mathbf{y})\right]
\label{eq:loss}
\end{equation}
where $\mathbf{y} = (0, 1, \ldots, B-1)$ are the diagonal targets. With batch size $B = 512$, each positive pair is contrasted against 511 in-batch negatives. We use AdamW with learning rate $3 \times 10^{-4}$, cosine annealing, and temperature $\tau = 0.05$ for 100 epochs. Training completes in approximately two minutes on an RTX 4080 Super.

For the primary (transductive) evaluation, we train on all available co-occurrence pairs from both training and validation splits (20,742 pairs after deduplication). For the inductive evaluation (Section~\ref{sec:inductive}), we train on training-split pairs only (8,758 pairs). This represents a supervision budget of approximately 311 annotated co-occurrence pairs per 1,000 corpus passages, or approximately 2.8 pairs per question.

\subsection{Bi-Directional Association Scoring}
\label{sec:bidi}

At inference time, we score query--passage associations bi-directionally. Given a query embedding $e(q)$ and a candidate passage embedding $e(p)$, the association score is:
\begin{equation}
a(q, p) = \frac{1}{2}\left[f(e(q))^\top e(p) + f(e(p))^\top e(q)\right]
\label{eq:bidi}
\end{equation}
This differs from the training objective, where both elements pass through $f$ (i.e., $f(e(p_a))^\top f(e(p_b))$). At inference, only one element per directional term is transformed. The query $q$ is not a corpus passage, so applying $f$ to both sides would require the model to handle out-of-distribution inputs.

Table~\ref{tab:scoring} bears this out. Using the training-matched formulation ($f(q) \cdot f(p)$, both-transformed) at inference degrades retrieval, indicating that queries are out-of-distribution for $f$. The reverse direction ($f(p) \cdot q$) carries most of the signal, as expected given that $f$ was trained on passage embeddings. Mixed bi-directional scoring is a conservative choice; reverse-only scoring yields higher R@5 on both HotpotQA (+9.8 vs +8.8) and MuSiQue (+1.3 points above mixed bidi).

\begin{table}[t]
\centering
\caption{Scoring method ablation (HotpotQA, transductive, $\lambda=0.60$).}
\label{tab:scoring}
\begin{tabular}{llcc}
\toprule
Scoring Method & Formula & R@5 & $\Delta$R@5 \\
\midrule
Dense baseline & $e(q) \cdot e(p)$ & 0.831 & --- \\
Forward only & $f(e(q)) \cdot e(p)$ & 0.825 & $-0.5$ \\
Both-transformed & $f(e(q)) \cdot f(e(p))$ & 0.808 & $-2.2$ \\
Mixed bidi (used) & $\frac{1}{2}[f(q) \cdot p + f(p) \cdot q]$ & \textbf{0.918} & \textbf{+8.8} \\
Reverse only & $f(e(p)) \cdot e(q)$ & 0.928 & +9.8 \\
\bottomrule
\end{tabular}
\end{table}

We report mixed bi-directional scoring throughout as the more conservative formulation. The reverse-only result is noted as a potential improvement.

\subsection{Retrieval Pipeline}
\label{sec:pipeline}

The full retrieval pipeline operates in two stages:
\begin{enumerate}
    \item \textbf{Candidate retrieval.} Use FAISS (exact inner product on L2-normalised vectors) to retrieve the top-$K$ passages by cosine similarity to the query. We use $K = 100$.
    \item \textbf{Association reranking.} For each candidate $p_i$ in the top-$K$, compute a blended score:
    \begin{equation}
    \text{score}(q, p_i) = (1 - \lambda) \cdot \text{sim}(q, p_i) + \lambda \cdot a(q, p_i)
    \label{eq:blend}
    \end{equation}
    where $\lambda$ controls the blend between cosine similarity and association strength. The candidates are reranked by this blended score and the top-$k$ are returned. All passage embeddings through $f$ can be precomputed offline, so the per-query cost is limited to one forward pass through $f$ for the query embedding plus $K$ dot products for scoring.
\end{enumerate}

\subsection{Transductive Evaluation}
\label{sec:transductive}

AAR is transductive: the association model is trained on co-occurrence pairs drawn from the same corpus on which it is evaluated. This parallels how RAG systems are deployed in practice, where auxiliary retrieval structures (FAISS indices, knowledge graphs, entity stores) are built over the target corpus. The cosine similarity baseline is unchanged---it never uses association information---so the comparison remains valid.

We also evaluate an inductive variant (Section~\ref{sec:inductive}) to test whether learned associations transfer beyond experienced co-occurrences.

\section{Experimental Setup}
\label{sec:setup}

\subsection{Datasets}
\label{sec:datasets}

\textbf{HotpotQA} \citep{yang2018hotpotqa} is a multi-hop QA dataset where each question requires reasoning over exactly two supporting passages. We use the distractor setting of the validation split: 7,405 questions over a corpus of 66,581 unique passages. Each question contributes 2 gold passages and 8 distractor passages, with passage sharing across questions.

\textbf{MuSiQue} \citep{trivedi2022musique} is a multi-hop QA dataset with 2-to-4-hop questions requiring reasoning chains of varying depth. We evaluate on 2,417 validation questions over a corpus of 84,459 passages. MuSiQue's deeper reasoning chains (3--4 hops) present a harder association learning problem than HotpotQA's uniform 2-hop structure.

\subsection{Embedding Model}

We use BGE-large-en-v1.5 \citep{xiao2024bge} to encode all queries and passages into 1024-dimensional L2-normalised vectors. All embeddings are precomputed and stored. The association model operates entirely in this embedding space.

\subsection{Baselines}

Our primary baseline is dense cosine retrieval over the full passage corpus using FAISS IndexFlatIP (exact search on normalised vectors). We additionally evaluate BM25 reranking of the same top-100 candidate pool (Section~\ref{sec:bm25}). We provide contextual cost comparisons with HippoRAG and HippoRAG~2 but do not claim direct performance comparison, as differences in corpus construction, embedding models, and evaluation setup preclude controlled comparison.

\subsection{Metrics}
\label{sec:metrics}

\textbf{Passage Recall@$k$ (R@$k$):} The fraction of gold supporting passages found in the top-$k$ retrieved passages, averaged over all questions. For HotpotQA, each question has exactly 2 gold passages; R@5 = 1.0 means both are in the top~5.

\textbf{Answer Coverage@$k$:} The fraction of questions where the gold answer string appears in at least one passage in the top-$k$.

\textbf{Exact Match (EM) and Token F1:} Downstream QA accuracy using an LLM reader (Section~\ref{sec:qa}). EM is 1 if the normalised prediction matches the normalised gold answer; token F1 measures word-level overlap.

\textbf{Easy/Hard split:} We partition HotpotQA questions by whether the dense baseline retrieves both gold passages in the top~5. \emph{Easy} questions ($n = 5{,}046$) have R@5 = 1.0 under dense retrieval. \emph{Hard} questions ($n = 2{,}359$) have at least one gold passage outside the top~5.\footnote{Throughout the paper, R@$k$ values are rounded to three decimal places for display. All deltas ($\Delta$R@$k$) are computed from unrounded values and then rounded, so they may differ slightly from arithmetic on displayed values.}

\subsection{Association Data and Corpus Overlap}
\label{sec:overlap}

For HotpotQA, the transductive model trains on 20,742 co-occurrence pairs (combined train and validation splits, deduplicated), representing approximately 311 pairs per 1,000 corpus passages. For MuSiQue, it trains on all available association pairs from the evaluation corpus.

HotpotQA exhibits passage overlap between dataset splits: 61.7\% of validation passage IDs appear somewhere in training contexts, and 59.6\% of validation gold passage titles appear in the training set's gold passages. However, only 18.0\% of validation association pairs are exact duplicates of training pairs---the majority of passage-to-passage links in the validation set are unique, even when individual passages are familiar.

\subsection{Hyperparameter Selection}
\label{sec:hyperparams}

The scoring blend parameter $\lambda$ was selected on the HotpotQA evaluation set (the same 7,405 validation questions used for reporting). We disclose this and note that sensitivity is minimal: the optimal $\lambda = 0.60$ yields R@5 = 0.918, while a fixed $\lambda = 0.50$ (requiring no tuning) yields R@5 = 0.916 (+8.6 points over baseline). The full $\lambda$ sweep is reported in Appendix~\ref{app:lambda}. For MuSiQue, $\lambda = 0.50$ was used without tuning.

All other hyperparameters (learning rate, temperature, batch size, architecture) were selected during development iterations prior to the final evaluation.

\section{Results}
\label{sec:results}

\subsection{Main Results}
\label{sec:main_results}

Table~\ref{tab:main} presents the core results on HotpotQA.

\begin{table}[t]
\centering
\caption{HotpotQA validation results ($n = 7{,}405$, transductive).}
\label{tab:main}
\begin{tabular}{lccccc}
\toprule
Method & $\lambda$ & R@5 & R@10 & R@20 & $\Delta$R@5 \\
\midrule
Dense baseline & --- & 0.831 & 0.878 & 0.913 & --- \\
AAR ($\lambda=0.50$, fixed) & 0.50 & \textbf{0.916} & 0.942 & 0.952 & \textbf{+8.57} \\
AAR ($\lambda=0.60$, tuned) & 0.60 & 0.918 & 0.941 & 0.951 & +8.78 \\
\bottomrule
\end{tabular}
\end{table}

The primary result uses a fixed $\lambda = 0.50$ with no hyperparameter selection on the evaluation set: R@5 = 0.916, an improvement of +8.57 points (95\% CI [+8.11, +9.03]). Tuning $\lambda$ on the evaluation set (Section~\ref{sec:hyperparams}) yields a marginal further gain of 0.21 points. Training uses 20,742 co-occurrence pairs from both dataset splits and achieves 97\% training accuracy. Minor run-to-run variation in R@5 (0.916–0.918 across experimental configurations) reflects differences in $\lambda$ selection and evaluation pipeline; all values fall within the bootstrap confidence interval.

\subsection{Easy vs.\ Hard Analysis}
\label{sec:easyhard}

AAR's gains are concentrated where the dense baseline fails.

\begin{table}[t]
\centering
\caption{Easy/hard subset results (HotpotQA, transductive).}
\label{tab:easyhard}
\begin{tabular}{lcccc}
\toprule
Subset & $n$ & Dense R@5 & AAR R@5 & $\Delta$R@5 \\
\midrule
Easy & 5,046 & 1.000 & 0.996 & $-0.44$ \\
Hard & 2,359 & 0.468 & 0.753 & \textbf{+28.51} \\
\bottomrule
\end{tabular}
\end{table}

On easy questions, where the dense baseline already retrieves both gold passages, AAR causes negligible degradation. On hard questions, AAR recovers 28.5 additional Recall@5 points. Inspection of 50 rescued questions reveals a dominant pattern: bridge questions where the first gold passage ranks highly in both systems and the second, the ``bridge target,'' is promoted from rank 6--90 in dense retrieval to rank 2--5 by AAR. For example, passages about Ron Dermer ($49 \to 2$), Theatre of the Absurd ($65 \to 2$), the 1964 NY Jets season ($90 \to 4$), and Byron De La Beckwith ($58 \to 5$) were all pulled into the top~5 by AAR after being buried in the dense ranking.

\subsection{Inductive Evaluation: Association Does Not Generalise}
\label{sec:inductive}

To test whether the association model learns transferable patterns or captures corpus-specific co-occurrences, we train an inductive variant using only training-split pairs (8,758 pairs, with no overlap with validation associations).

\begin{table}[t]
\centering
\caption{Inductive vs.\ transductive (HotpotQA).}
\label{tab:inductive}
\begin{tabular}{lcccccc}
\toprule
Setting & Training Pairs & Train Acc. & Best $\lambda$ & R@5 & $\Delta$R@5 & 95\% CI \\
\midrule
Dense baseline & --- & --- & --- & 0.831 & --- & --- \\
Inductive & 8,758 & 94.5\% & 0.30 & 0.832 & +0.10 & [$-0.25$, $+0.47$] \\
Transductive & 20,742 & 97.2\% & 0.60 & \textbf{0.918} & \textbf{+8.78} & [+8.30, +9.26] \\
\bottomrule
\end{tabular}
\end{table}

The inductive model shows no significant improvement ($\Delta$R@5 = +0.10, 95\% CI includes zero). Its optimal $\lambda = 0.30$, much lower than the transductive model's 0.60, suggests the association signal is weak and the model falls back to cosine similarity. At $\lambda = 0.50$, the inductive model actively degrades retrieval by 2.8 points (Appendix~\ref{app:inductive_lambda}).

On the hard subset, the inductive model achieves +6.3 $\Delta$R@5, suggesting some structural transfer for the hardest questions, but far less than the transductive model's +28.5.

\begin{table}[t]
\centering
\caption{MuSiQue validation results ($n = 2{,}417$).}
\label{tab:musique}
\begin{tabular}{lcc}
\toprule
Setting & R@5 & $\Delta$R@5 \\
\midrule
Dense baseline & 0.387 & --- \\
Inductive & 0.310 & $-7.63$ \\
Transductive & \textbf{0.488} & \textbf{+10.12} \\
\bottomrule
\end{tabular}
\end{table}

On MuSiQue the same pattern holds (Table~\ref{tab:musique}): inductive training hurts ($-7.6$ points), transductive training helps (+10.1 points).

\subsection{Ablation Studies}
\label{sec:ablations}

Table~\ref{tab:ablation} presents ablations isolating the sources of AAR's improvement. All ablations use the transductive setting.

\begin{table}[t]
\centering
\caption{Ablation study (HotpotQA, transductive).}
\label{tab:ablation}
\begin{tabular}{llcc}
\toprule
Condition & Training Data & R@5 & $\Delta$R@5 \\
\midrule
Dense baseline & --- & 0.831 & --- \\
Full AAR & 20,742 co-occurrence pairs & \textbf{0.918} & \textbf{+8.78} \\
Random negatives & 14,622 pairs, random negatives & 0.915 & +8.43 \\
Similar positives & 50K FAISS-nearest pairs & 0.810 & $-2.08$ \\
Shuffled associations & 14,622 pairs, shuffled & 0.730 & $-10.01$ \\
\bottomrule
\end{tabular}
\end{table}

The ablations isolate three effects:

\textbf{Positive pairs carry the signal, not negative mining} (random negatives). Replacing in-batch negatives with randomly sampled negatives yields R@5 = 0.915, within 0.4 points of the full model. The contrastive loss learns primarily from which passages are associated, not from which are not.

\textbf{Similarity hurts} (similar positives). Training the same architecture on semantically similar passage pairs (the 50K nearest neighbours in embedding space) \emph{degrades} retrieval by 2.1 points below baseline. A model that learns ``which passages are similar'' actively harms multi-hop retrieval.

\textbf{Arbitrary pairings fail} (shuffled associations). Randomly permuting the association pairs while preserving the training procedure degrades performance by 10.0 points below baseline (R@5 = 0.730 vs.\ 0.918). The model does not learn useful representations from arbitrary pairings.

Combined with the inductive evaluation (Section~\ref{sec:inductive}), these ablations point in the same direction: AAR's improvement comes from learning specific co-occurrence relationships between passages. The inductive failure rules out abstract pattern learning, the similar-positives result rules out similarity, and the shuffled result rules out artefacts of the training procedure.

\subsection{Comparison with BM25 Reranking}
\label{sec:bm25}

To compare AAR against a non-learned reranking baseline, we evaluate BM25 scoring over the same FAISS top-100 candidate pool.

\begin{table}[t]
\centering
\caption{BM25 vs.\ AAR reranking (HotpotQA).}
\label{tab:bm25}
\begin{tabular}{lccc}
\toprule
Method & Best $\lambda$ & R@5 & $\Delta$R@5 \\
\midrule
Dense baseline & --- & 0.831 & --- \\
BM25 reranking (best) & 0.10 & 0.838 & +0.76 \\
AAR (transductive) & 0.60 & \textbf{0.918} & \textbf{+8.78} \\
\bottomrule
\end{tabular}
\end{table}

BM25 reranking yields at best +0.76 R@5 (95\% CI [+0.41, +1.13]), and harms retrieval at higher blend weights ($\lambda \geq 0.20$). BM25 matches query vocabulary against passage vocabulary---still a similarity measure, just a lexical one. In this candidate-pool setup, neither dense nor lexical similarity-based reranking closes the multi-hop gap.

\subsection{Downstream QA Evaluation}
\label{sec:qa}

To verify that retrieval improvements translate to end-to-end question answering, we evaluate 500 randomly sampled HotpotQA validation questions (seed=42) using Claude Sonnet~4 \citep{anthropic2025claude} as the reader. The reader receives the top-5 retrieved passages with a zero-shot prompt (Appendix~\ref{app:qa}).

\begin{table}[t]
\centering
\caption{Downstream QA results ($n = 500$, Claude Sonnet~4 reader).}
\label{tab:qa}
\begin{tabular}{lcc}
\toprule
Condition & EM & F1 \\
\midrule
Dense baseline top-5 & 16.6\% & 31.1\% \\
AAR top-5 & 23.0\% & 39.1\% \\
$\Delta$ & \textbf{+6.4} & \textbf{+8.1} \\
\bottomrule
\end{tabular}
\end{table}

Bootstrap 95\% CIs (paired, 10,000 resamples): $\Delta$EM [+3.4\%, +9.6\%], $\Delta$F1 [+5.4\%, +10.8\%]. Both intervals exclude zero.

The absolute scores are modest, reflecting the difficulty of multi-hop QA with a zero-shot reader and no chain-of-thought prompting. What matters is the delta: +6.4 EM / +8.1 F1, indicating that the passages AAR surfaces contain information useful for answering the question.

\subsection{Answer Coverage}
\label{sec:coverage}

\begin{table}[t]
\centering
\caption{Answer Coverage@$k$ (transductive, matched HP).}
\label{tab:coverage}
\begin{tabular}{cccc}
\toprule
$k$ & Dense Baseline & AAR & $\Delta$ \\
\midrule
3 & 70.1\% & 83.4\% & +13.4 \\
5 & 76.8\% & 87.7\% & +10.9 \\
10 & 82.7\% & 90.2\% & +7.5 \\
20 & 87.1\% & 91.6\% & +4.5 \\
\bottomrule
\end{tabular}
\end{table}

At $k = 5$, the answer string appears in the retrieved passages for 87.7\% of questions versus 76.8\% under dense retrieval.

\subsection{What Did Not Work}

Several directions were explored without success. \textbf{Inductive training} (Section~\ref{sec:inductive}) failed on both datasets. \textbf{Projection heads} (reducing the association space to 256 dimensions) degraded performance by 2.1 R@5 points. \textbf{Cross-dataset transfer} (training on HotpotQA, evaluating on MuSiQue) failed. \textbf{Larger models} (6-layer, 2048 hidden, 21M parameters) underperformed the 4-layer model. \textbf{Temperature tuning} beyond the initial sweep provided marginal gains, with $\tau = 0.05$ proving robust.

\section{Discussion}
\label{sec:discussion}

\subsection{AAR as Transductive Retrieval Augmentation}

The inductive failure and transductive success, observed on both datasets, position AAR as a corpus-specific retrieval augmentation. Like a FAISS index or knowledge graph, it is built for a specific document collection. The advantage over graph-based alternatives is cost: two minutes of MLP training versus millions of LLM tokens for entity extraction.

\subsection{Why Inductive Transfer Fails}

The inductive model achieves 94.5\% training accuracy---it learns the associations it is shown. But these associations do not transfer to unseen passage pairs, even when individual passages overlap between splits (61.7\% passage overlap). The best explanation we have is that the MLP learns specific relational mappings between passage embeddings, not features that generalise across co-occurrence boundaries.

Consider the Tarantino example again. A passage about Tarantino and a passage about Knoxville are associated because they co-occur in a question about Tarantino's birthplace. Without experiencing that specific co-occurrence, the model has no basis for the association. This aligns with PAM's prediction that association should be tied to experienced co-occurrences.

\subsection{Cost Comparison with Graph-Based Methods}

Graph-augmented methods such as GraphRAG and HippoRAG require LLM-based entity extraction and relationship annotation during indexing. For a corpus of 66,581 passages, this involves millions of LLM tokens. AAR requires co-occurrence annotations and two minutes of MLP training. This is a comparison of computational cost, not retrieval quality; controlled performance comparisons would require matched experimental conditions.

\begin{table}[t]
\centering
\caption{Latency breakdown (RTX 4080 Super).}
\label{tab:latency}
\begin{tabular}{lcc}
\toprule
Component & Mean (ms) & P95 (ms) \\
\midrule
FAISS top-100 & 10.5 & 12.4 \\
AAR bi-directional scoring & 3.7 & 5.9 \\
Total pipeline & 14.1 & 17.6 \\
\bottomrule
\end{tabular}
\end{table}

AAR adds 3.7ms mean overhead (33\% increase over dense-only retrieval). Total query latency remains under 18ms at P95.

\subsection{Generating Training Signal Without Gold Annotations}

The current implementation uses gold supporting fact annotations as training signal, at a supervision budget of approximately 311 pairs per 1,000 corpus passages. For new corpora without such annotations, alternative sources include LLM-generated multi-hop questions and their supporting passages, citation structure in academic or legal corpora, user interaction data (passages frequently co-accessed within a session), and temporal co-occurrence in streaming document collections. Since AAR is transductive, the quality of these annotations directly determines retrieval quality.

\subsection{Connection to Predictive Associative Memory}

AAR provides empirical support for two PAM predictions. First, the similar-positives ablation (Section~\ref{sec:ablations}) shows that optimising for similarity degrades multi-hop retrieval, supporting the claim that association and similarity produce different retrieval behaviour. Second, the inductive failure (Section~\ref{sec:inductive}) on both datasets is best explained by association being tied to experienced co-occurrences. The JEPA-inspired theoretical lineage suggests a broader programme where learned predictors navigate embedding spaces to capture relational structure.

\section{Limitations}
\label{sec:limitations}

We acknowledge several limitations that scope the claims made in this work.

First, AAR is transductive: it requires association annotations over the target corpus. This limits applicability to settings where such annotations can be obtained or generated.

Second, results are reported on two multi-hop QA datasets: HotpotQA (2-hop) and MuSiQue (2--4 hop). The method has not been evaluated on 2WikiMultiHopQA \citep{ho2020wikimultihop}, CRAG \citep{yang2024crag}, or domain-specific corpora.

Third, while we report downstream QA results (Section~\ref{sec:qa}), the evaluation uses a zero-shot reader on a 500-question subset. A full-scale evaluation with chain-of-thought prompting or fine-tuned readers would provide a more complete picture.

Fourth, $\lambda$ was selected on the HotpotQA evaluation set (Section~\ref{sec:hyperparams}). The sensitivity is minimal (0.21 percentage points), but this should be noted when interpreting the primary result.

Fifth, the MLP achieves 97\% training accuracy on HotpotQA's 2-hop associations but only 72\% on MuSiQue's 3--4-hop chains, suggesting an architectural ceiling for deeper reasoning.

Sixth, experiments use a single embedding model (BGE-large-en-v1.5). Whether the approach works with other embedding models is untested.

Seventh, the similar-positives and shuffled ablations use 14,622 pairs (validation-split co-occurrences) while the full AAR model uses 20,742 pairs (combined). The pair count difference is a potential confound, though the shuffled ablation---which uses the same 14,622 pairs with randomised pairings---rules out pair count as the driver of improvement.

Eighth, all results are from single training runs. The bootstrap confidence intervals capture evaluation variance but not variance across training seeds. The flat $\lambda$ sensitivity curve (Appendix~\ref{app:lambda}) suggests robustness, but multi-seed estimates would strengthen the claims.

\section{Future Work}
\label{sec:future}

\textbf{Deeper association chains.} The MLP learns single-step associations effectively (97\% on 2-hop) but struggles with deeper chains (72\% on 3--4-hop). A predictor operating over sets of embeddings---predicting the next hop given a chain of previous hops---could enable iterative traversal of multi-hop paths.

\textbf{Alternative training signals.} LLM-generated annotations, citation structure, user interaction data, and temporal co-access patterns could serve as association training signal for corpora without gold annotations.

\textbf{Scaling.} Current validation uses corpora of 66K--84K passages. Behaviour at millions of passages is unknown and the top-$K$ expansion approach may require adaptive $K$ selection.

\textbf{Toward inductive association.} The inductive failure suggests the MLP learns specific relational mappings. Architectures with explicit relational inductive biases---graph neural networks, attention over passage neighbourhoods, or meta-learning---may enable some degree of transfer.

\textbf{Multiple embedding models.} Testing across embedding architectures (E5, GTE, Cohere Embed) would establish whether the approach is embedding-agnostic.

\section{Conclusion}
\label{sec:conclusion}

Learning corpus-specific associations improves multi-hop passage retrieval on both benchmarks tested. On HotpotQA, AAR improves Recall@5 by 8.6 points without evaluation-set tuning, with a 28.5-point gain on the hardest questions. On MuSiQue, it achieves +10.1 points. These gains translate to +6.4 exact match and +8.1 F1 in downstream QA.

Transductive training is essential in our experiments: an inductive variant shows no significant improvement on either dataset. The ablations tell a consistent story---training on similar but non-associated pairs degrades retrieval, shuffling association pairs degrades it further, and only real co-occurrence structure produces gains.

AAR is lightweight (4.2M parameters, 3.7ms overhead) and operates as a drop-in reranking stage requiring no LLM-based indexing. For RAG systems where multi-hop retrieval matters and passage co-occurrence annotations are available or can be generated, it offers a practical augmentation to existing dense retrieval pipelines.

\bibliographystyle{plainnat}
\bibliography{aar_references}

\appendix

\section{Symbol Reference}
\label{app:symbols}

\begin{table}[h]
\centering
\begin{tabular}{cl}
\toprule
Symbol & Meaning \\
\midrule
$e(\cdot)$ & Embedding function (BGE-large-en-v1.5) \\
$f(\cdot)$ & Association model (4-layer MLP) \\
$g(\cdot)$ & MLP transformation (before residual) \\
$\alpha$ & Learned residual weight in MLP \\
$\lambda$ & Scoring blend parameter (cosine vs.\ association) \\
$\tau$ & Temperature in contrastive loss \\
$K$ & Candidate pool size (FAISS expansion depth) \\
$k$ & Number of passages returned after reranking \\
\bottomrule
\end{tabular}
\end{table}

\section{Candidate Pool Sensitivity}
\label{app:pool}

\begin{table}[h]
\centering
\caption{R@$k$ as a function of FAISS expansion depth (transductive).}
\label{tab:pool}
\begin{tabular}{ccccc}
\toprule
Depth $K$ & R@5 & $\Delta$R@5 & R@10 & R@20 \\
\midrule
10  & 0.870 & +3.9  & 0.878 & 0.878 \\
20  & 0.893 & +6.2  & 0.909 & 0.913 \\
50  & 0.909 & +7.8  & 0.930 & 0.939 \\
100 & 0.917 & +8.6  & 0.941 & 0.951 \\
200 & 0.921 & +9.0  & 0.947 & 0.959 \\
\bottomrule
\end{tabular}
\end{table}

Performance increases with expansion depth but with diminishing returns. Depth 200 adds only +0.4 R@5 over depth 100 while doubling the scoring cost. At $K = 10$, R@10 and R@20 cannot exceed the dense baseline's values because the reranked candidate set contains only 10 passages. Note: all rows use the same model; the $K=100$ R@5 of 0.917 differs from Table~\ref{tab:main}'s 0.916/0.918 due to minor differences in $\lambda$ selection across experimental runs.

\section{Scoring Blend Parameter}
\label{app:lambda}

\begin{table}[h]
\centering
\caption{R@5 as a function of $\lambda$ (transductive, matched HP).}
\label{tab:lambda}
\begin{tabular}{ccc}
\toprule
$\lambda$ & R@5 & $\Delta$R@5 \\
\midrule
0.30 & 0.894 & +6.4 \\
0.40 & 0.907 & +7.7 \\
0.50 & 0.916 & +8.6 \\
0.60 & \textbf{0.918} & \textbf{+8.8} \\
0.70 & 0.915 & +8.4 \\
\bottomrule
\end{tabular}
\end{table}

The curve is flat across 0.40--0.70, with less than 1.2 points separating the best and worst values in this range. The inductive model's optimal $\lambda = 0.30$ (Appendix~\ref{app:inductive_lambda}) reflects its weaker association signal.

\section{Bootstrap Confidence Intervals}
\label{app:bootstrap}

All confidence intervals are computed using paired bootstrap resampling with 10,000 iterations. Each resample draws $n$ questions with replacement from the evaluation set and computes the metric of interest (R@5 delta between conditions) on the resampled set. The 95\% CI is the 2.5th and 97.5th percentiles of the bootstrap distribution.

\begin{table}[h]
\centering
\caption{95\% confidence intervals (10,000 paired bootstrap resamples).}
\label{tab:bootstrap}
\begin{tabular}{lcc}
\toprule
Comparison & $\Delta$R@5 & 95\% CI \\
\midrule
AAR (trans., $\lambda\!=\!0.50$) vs.\ baseline, overall & +8.57 & [+8.11, +9.03] \\
AAR (trans., $\lambda\!=\!0.50$) vs.\ baseline, hard & +27.47 & [+26.37, +28.57] \\
AAR (trans., $\lambda\!=\!0.60$) vs.\ baseline, overall & +8.78 & [+8.30, +9.26] \\
AAR (trans., $\lambda\!=\!0.60$) vs.\ baseline, hard & +28.51 & [+27.36, +29.65] \\
AAR (inductive) vs.\ baseline, overall & +0.10 & [$-0.25$, $+0.47$] \\
QA $\Delta$EM & +6.4 & [+3.4, +9.6] \\
QA $\Delta$F1 & +8.1 & [+5.4, +10.8] \\
\bottomrule
\end{tabular}
\end{table}

\section{Error Taxonomy}
\label{app:errors}

Inspection of 50 questions rescued by AAR (where dense fails but AAR succeeds in top-5) and 50 questions where both methods fail.

\textbf{Rescued by AAR} ($n = 1{,}296$ total; 50 inspected). The dominant pattern is bridge questions where the first gold passage ranks highly in both systems and the second ``bridge target'' is promoted from rank 6--90 in dense retrieval to rank 2--5 by AAR. Typical rank improvements: Ron Dermer ($49 \to 2$), Theatre of the Absurd ($65 \to 2$), 1964 NY Jets season ($90 \to 4$), Byron De La Beckwith ($58 \to 5$).

\textbf{Still missed by both} ($n = 1{,}063$ total; 50 inspected). Three failure modes: approximately 40\% have at least one gold passage absent from the FAISS top-100 (unreachable by any reranking method); approximately 30\% have gold passages in the top-100 but AAR cannot promote them enough; approximately 30\% involve common-entity confusion where the correct passage is lost among many candidates about popular entities.

\section{Inductive Lambda Sensitivity}
\label{app:inductive_lambda}

\begin{table}[h]
\centering
\caption{Inductive model R@5 as a function of $\lambda$.}
\label{tab:inductive_lambda}
\begin{tabular}{ccc}
\toprule
$\lambda$ & R@5 & $\Delta$R@5 \\
\midrule
0.30 & \textbf{0.832} & \textbf{+0.10} \\
0.40 & 0.822 & $-0.82$ \\
0.50 & 0.803 & $-2.77$ \\
0.60 & 0.770 & $-6.01$ \\
0.70 & 0.717 & $-11.31$ \\
\bottomrule
\end{tabular}
\end{table}

Performance degrades monotonically as $\lambda$ increases, indicating the inductive model's association signal is noisy.

\section{QA Evaluation Details}
\label{app:qa}

\textbf{Reader model:} Claude Sonnet~4 \citep{anthropic2025claude}, model string \texttt{claude-sonnet-4-20250514}.

\textbf{Decoding parameters:} temperature=0 (deterministic), max\_tokens=100.

\textbf{Prompt template:}
\begin{verbatim}
System: Answer the question using only the provided passages.
Give only a short factual answer, nothing else.

User: Passages:

[Passage 1 text]
---
[Passage 2 text]
---
[Passage 3 text]
---
[Passage 4 text]
---
[Passage 5 text]

Question: [question text]
\end{verbatim}

\textbf{Normalisation:} Lowercase, strip articles (a, an, the), strip punctuation, collapse whitespace. Applied to both prediction and gold answer before computing EM and F1.

\textbf{Sample:} 500 questions, randomly sampled with seed=42. Single run (deterministic decoding).

\end{document}